\documentclass{article}

\usepackage{spconf,amsmath,epsfig}
\usepackage{amsthm}
\usepackage{float}
\newfloat{algorithm}{t}{lop}
\floatname{algorithm}{Algorithm}
\usepackage[noend]{algpseudocode}

\usepackage{graphicx}
\usepackage{epstopdf}
\usepackage{tikz}

\usepackage{verbatim}

\usepackage{url}
\usepackage{dblfloatfix}

\DeclareGraphicsExtensions{.ps}
\begin{document}
%

\title{Who Ordered This?: Exploiting Implicit User Tag Order Preferences for Personalized Image Tagging}
\name{Amandianeze O. Nwana and Tsuhan Chen}

\address{School of Electrical and Computer Engineering, Cornell University, Ithaca, NY 14850, U.S.A.}

\maketitle

\begin{abstract}
What makes a person pick certain tags over others when tagging an 
image? Does the order that a person presents tags for a given image follow
an implicit bias that is personal? Can these biases be used to improve existing
automated image tagging systems? We show that tag ordering, which has been 
largely overlooked by the image tagging community, is an important cue in 
understanding user tagging behavior and can be used to improve auto-tagging
systems. Inspired by the assumption that people order their tags, we propose a
new way of measuring tag preferences, and also propose a new personalized 
tagging objective function that explicitly considers a user's preferred tag
orderings. We also provide a (partially) greedy algorithm that produces good
solutions to our new objective and under certain conditions produces an optimal
solution. We validate our method on a subset of Flickr images that spans 5000 
users, over 5200 tags, and over 90,000 images. Our experiments show that 
exploiting personalized tag orders improves the average performance of 
state-of-art approaches both on per-image and per-user bases.

\end{abstract}

%



\section{Introduction}
\label{sect:intro}

Nearly every work on image tagging to date has treated the tags that accompany
an image as a bag-of-words with no inherent ordering, most especially works 
which use the nearest neighbor approach for tagging. Some recent work
\cite{nwana:15}, has shown that this bag-of-words assumption is actually not 
the reality. They show that if a user is asked to tag an image with a fixed set
of tags multiple times, the order of the tags they choose tends to stay the 
same rather than random.
More so, a recent design change on Flickr~\cite{web:flickr} that reversed the
users' tag input ordering created a backlash from the community that caused
Flickr to revert back to the original design with the following apology:
``We thought it would be more intuitive for newest tags to appear at the top,
so that when you add a tag and refresh, it's in the place you expect it. 
However, thanks to your points about meticulous tag ordering, we've decided we
should leave it as is. You should now see your tags in the correct order''
\cite{web:flickr_order}.

These insights lead us to believe that tag ordering is inherently personal, and
should be incorportated into tagging systems. Hence, we propose a new model
that seeks to exploit the history of a user's tag order to improve personalized
image tagging. We also compare the same model which instead treats tag order as
a global phenomenon, and show that indeed such a model underperforms the 
personalized ordering model significantly.


\subsection{Related Work}
\label{sec:related}

There has been a lot of work in the area of automatic image tagging
\cite{li:11, weston:11, lindstaedt:08, mcauley:12, sawant:10, 
sigurbjornsson:08,stone:08, belem:11}, much of which has treated tagging as
more of a labelling problem, thereby implicitly imposing that tagging is a 
global task rather than a user-specific task. Most notably the work by Li et
al.~\cite{li:11} has tried to treat the tagging problem from the point of
view of personalization by using a cross-entropy method to decide how to 
weight different image tagging functions for each user.

There has been work seeking to tag based on object importance, 
but there the focus has been an explicitly categorical approach to 
importance by measuring properties of objects in the image (eg, size, salience)
to estimate their relative importances \cite{berg:12}. These object-property
based approaches to importance typically ignore particular user preferences, and
treat importance as a global phenomenon \cite{berg:12, spain:11}. 

Nearest neighbor approaches are usually more 
common than their explicit classification counterparts in tagging because one
does not have to learn how to recognize or detect specific objects
in the image, which is not scalable, nor are all concepts one
would like to describe an image visual
\cite{smeulders:00, schreiber:01, weston:11}. 
In many cases, an initial set of tags for a given query
image is provided and the aim is to refine or propagate these tags through some, 
similarity measure (image, user, or tag co-similarity) \cite{lindstaedt:08, 
rattenbury:07, schmitz:06}. Whereas their goal is to create global tags from
other tags, our goal is to find the words most suitable to describe an
image for specific individuals.
There has also been some work that looks at the recommendation perspective 
\cite{adomavicius:05, koren:09,rendle:09,
rendle:10, pan:13, peng:14, sahoo:12, sen:09, song:08},
with~\cite{peng:14} having an explicit goal of personalized tagging.

In this work, we employ the nearest neighbor approach to tagging because we would like to
use a relatively large tagging vocabulary, which introduces scalability issues
with the explicit/classification approaches, and we also do not want to restrict
our vocabulary to words that describe concrete visual concepts. Instead we
would like to have more abstract tags because we are primarily interested in 
personalized tagging. Finally, we also do not want to consider tag importances
solely on a global scale, but per user preference.  


\section{Model}
\label{sec:model}

In this section we discuss our model for the tagging behavior of users
and derive an objective function based on the model of tagging behavior. We
then discuss a practical representation of this model and how it relates to
other known problems.


\subsection{User Tagging Behavior}
\label{model:assumption}

The main assumption, following the claims from~\cite{nwana:15},
is that given an image, the order that 
users present tags to an online tagging system, say Flickr \cite{web:flickr}, 
implies an underlying preference order for those tags.  For example, if a user
presents the tags $B,\ A,\ C$ in that order, it would imply that for that image,
the user found it more preferable (or important) to mention $B$ before $A$,
$B$ before $C$, and $A$ before $C$. 

Our main idea is that with enough observed instances of pairs of tags mentioned 
together by a particular user, we can estimate these implicit biases, and in turn 
improve the task of automatic image tagging for new images on a personal level.


\subsection{Tagging Objective}
\label{model:objective}

Our main tagging goal is to output a list of tags for an image in order of
preference for the target user. To that end, given a set of pairwise 
tag-preferences (as probabilities) for a given user and a query image, we want
to find the ordering of candidate tags (generated by some baseline tag 
generator), for that query image, that is maximal. 
Details are provided in the supplementary appendix.

\subsubsection*{Parameter Estimation}
\label{model: parameter_estimation}

For the objective described above, we represent the tag-preferences
as probabilities. That is, $p_{ab}$ is the strength of preference for tag $a$ over
tag $b$. We estimate this quantity as the number of times $a$ occurs before $b$,
divided by the number of times they occur together, regardless of order. This
allows for anti-symmetry: $p_{ba} = 1 -p_{ab}$. In this representation, $p_{ab}=.5$
implies there is no preference. Also note the we only need to keep track of the strongest
preference (either $p_{ab}$, or $p_{ba}$, but not both).

\subsection{Maximizing Objective -- PrioritizedTopoSort}

To maximize our tagging objective, we came up with a modified version of the
topological sort algorithm. Since we can represent the preferences as directed
edges with weighs as the probabilities, we can use topological sort~\cite{toposort}
to produce an ordering that is faithful to these preferences.

To deal with cycles, which are assumed not to exist for the topological sort
algorithm to work, when we find a cycle, we delete the edge of least strength.
Under certain mild conditions, this is guaranteed to still be optimal. Also,
since there could be multiple correct topological sort orderings, in order to
break ties, we use a baseline ordering of the candidate tags 
(gotten from the baseline tag generator as mentioned in section
\ref{model:objective}), as we run the modified topological sort algorithm,
whenever there is an ambiguous ordering.

Details and proofs of claims made here are provided in the appendix.



\section{Experimental Setup}

In this section, we will discuss how we go about verifying our model, from 
the choice of the dataset, choice of baseline and choice of evaluation metric. 

\subsection{Dataset}

For this work we chose to work with the NUS-WIDE dataset \cite{nus-wide-civr09}.
The NUS-WIDE dataset is a subset of 269,648 images from Flickr. For each image in
the dataset, we know, via the Flickr API, the corresponding user that uploaded
that image, and the sequence of tags that user chose to annotate the image with. 
Since we are particularly concerned about personalization, we only select images
from this database which satisfy the following criteria: the users who uploaded
the image must have at least 6 images in this dataset, similar to the setting in
\cite{li:11}. This results in about 91,400 images from 5000 users. We split this dataset
into a training and test partition, by randomly assigning half of each users' images
to the training, and half to the test set. 
For each image we only retain the tags that occurred frequently enough across
the dataset, in order to make some sort of meaningful inference on the
tags. In this work, we made the design choice of working with tags that occurred
at least 50  times in the dataset. This results in a vocabulary of $5,326$ unique tags.

\subsubsection*{Image Similarity}
As we mentioned in section \ref{sec:related}, we go the nearest neighbor route for the
task of image tagging. To that end our image features are a
500-D bag of words based on SIFT descriptors \cite{lowe:04}, and we use the
euclidean distance between feature vectors to encode the notion of closeness or
similarity.

\subsection{Baseline}
\label{exp:baseline}

We take as our baseline the work of Li et al. \cite{li:11}, which
we considered the reproducable state of the art for \emph{personalized}
image tagging and evaluated their claims on the NUS-WIDE dataset as well. 
Their main idea
is that for a given tag, each user has 2 weighting variables, one weighting
how much to rank that tag according to its frequency independent of visual
content, and the other how much to rank the frequency of the tag according
to its frequency in visually similar images. We re-implemented
their method using the two tagging functions described in their paper (\textit{PersonalPreference}
\cite{sawant:10} function and \textit{Visual}\cite{li:10} function). For the visual 
features, we also use the 500-D bag of words based on SIFT descriptors for
consistency.

For a given image query $q$, we define the ordered set of tags returned by the
baseline method as $XE(q)$.

\subsection{Metrics}
\label{exp:metrics}
Since our assumption is that the order that a user tags an image important for
automatic tagging, our metric should take the groundtruth user order into account. 

More concretely, given an ordered set of tags,$\{t_1,
\cdots,t_k\}$, we define the relevance of each tag according to it's reciprocal
rank in the ordered set:
\[
	rel(t_i) = \frac{1}{rank^*(t_i)},\ \forall t_ i
\]
Tags not in the groundtruth ordered set have zero relevance.
More common metrics, such as precision, recall, and average precision, assume
that all tags are equally relevant, so we do not utilize these metrics in this
paper. Instead we use the more appropriate \emph{normalized discounted 
cumulative gain} (nDCG), a common metric used in evaluating
search engine results. For an ordered set $T = \{t_1,\cdots,t_k\}$, such that
$i < j \rightarrow t_i \succ t_j$, we define the DCG with respect to the 
ground-truth as:

\begin{equation}
	DCG(T) = rel(t_1) + \sum\limits_{t_i \in T, i\not=1} \frac{rel(t_i)}{\log_2(i)}
	\label{eqtn:dcg}
\end{equation}

This metric is called \emph{discounted} because the later we include a tag in our ranking,
the less gain we get from it (i.e. its relevance is discounted by the inverse of
the log of its position in the ranking, not the groundtruth). Note that this metric is
maximal when the most relevant items are listed first.

We also define for a given ranked list, $T$, its ideal ranking $\bar{\sigma}(T)$,
such that for $x,y \in T$, if $\bar{\sigma}^{-1}(x) < \bar{\sigma}^{-1}(y)$, 
then $x \succ y$ in the groundtruth (i.e., the ideal ranking is ranked from most 
relevant to least). Then the normalized DCG is defined as:
\[
	nDCG(T) = \dfrac{DCG(T)}{DCG(\bar{\sigma}(T))}
	\label{eqtn:ndcg}
\]

Note that the nDCG is maximal (equal to 1) when $T=\bar{\sigma}(T)$. We can also 
parameterize the DCG, and nDCG to calculate the $DCG@k$ and $nDCG@k$. That is, 
calculate the metrics evaluated only for the first $k$ entries of the ranked lists. Let
$T[:k]$ be the first k entries of $T$, then:
\[
  nDCG@k(T) = \dfrac{DCG(T[:k])}{DCG(\bar{\sigma}(T)[:k])}
\]


\section{Results}


\subsection{Task}

Given a query image $q$ from 
the test set, we find the $N$ nearest neighbors (visual similarity) from the 
training set, and given some initial rank $\tilde{T}$ on the set of tags used to 
describe these $N$ visually similar images, we find optimal ranking, 
$\sigma^*(\tilde{T})$, as described in section~\ref{model:objective}, 
using the preference probabilities derived
from the training set according to section~\ref{model: parameter_estimation}.
We set the baseline order of candidate tags as $\tilde{T} = XE(q)$, which is the
result of our baseline (and the current state-of-art for personalization)
given the query $q$, from section \ref{exp:baseline}.

\subsection{Design Decisions}
\label{res:design_dec}

In evaluating our approach, we made a few 
design decisions which are parameters in our model. The first
is the minimum number of times tags $a$ and $b$ occur together, which we
denote as $C$. If any pair of tags don't co-occur at least $C$ times,
we do not have a high confidence in the order bias that we estimate from
the occurrences because of noise and overfitting. The second which we denote
as $R$ is the minimum strength factor between a pair of tags. That is,
we ensure that, $p_{ab} \geq R.p_{ba}$, for $p_{ab}$ to be an edge to be
considered by the PrioritizedTopoSort. The purpose of these parameters
is to control for noise and overfitting to insufficient data.

\subsection{Evaluations}
 
 In this section, we describe the tests with which we evaluate our method.
 For the evaluation metrics discussed in section \ref{exp:metrics}, we
 run evaluations under the following settings:
 
 \begin{itemize}
 	\item Number of visually similar neighbors: $\{50,100\}$
 	\item Minimum co-occurrence, C: $\{5,10,30\}$
 	\item Minimum strength factor, R: $\{5,10,30\}$
 \end{itemize}
 
 We compare our performance (personalized pairwise preference), to the
 state-of-art~\cite{li:11}, and also to global pairwise preferences. Global
 pairwise preferences is similar to what we have proposed so far in this 
 paper with the exception of ignoring the users, so the images are treated
 as though they all come from the ``average user''. 
 
 We provide evaluation for our metrics on entire ranks, and also the top
 $k$ tags in the rank for $k \in \{1,5,10,15,30\}$. 
 We show the mean performance averaged both over each image (included in the 
 appendix), and also average across its mean performance per user since we 
 are interested in personalization.
 
 We also calculate the number of times our approach produces better rankings
 (in terms of NDCG), than the baseline, and the number of times a user prefers
 our approach on average over the baseline. We report these numbers in Table 
\ref{tab:duels}.

\subsection{Observations}

\begin{table}[t]
  \begin{center}
    \begin{tabular}{| l || c | c |}
    \hline
              & NN = 100  & NN = 50 \\
    \hline\hline
    Per User  & 5.5\%     & 4.8\%       \\
    \hline
    Per Image & 7.3\%     &  6.2\%  \\
    \hline
    \end{tabular}
  \end{center}
  \caption{Average NDCG percentage improvement over the baseline using pairwise
  personal preferences}
  \label{tab:avg_stats}
\end{table}

\begin{table}[t]
  \begin{center}
    \begin{tabular}{| l || c | c |}
    \hline
              & NN = 100  & NN = 50 \\
    \hline\hline
    Per User  & 1.11    & 1.10       \\
    \hline
    Per Image & 1.13     & 1.09  \\
    \hline
    \end{tabular}
  \end{center}
  \caption{Average number of times our approach is preferred to the baseline. Our
  approach is preferred about 10\% more of the time than the baseline (state-of-art).
  }
  \label{tab:duels}
\end{table}

\begin{figure*}[ht!]
 \centering
  \includegraphics[height=.27\textheight,width=.9\textwidth]{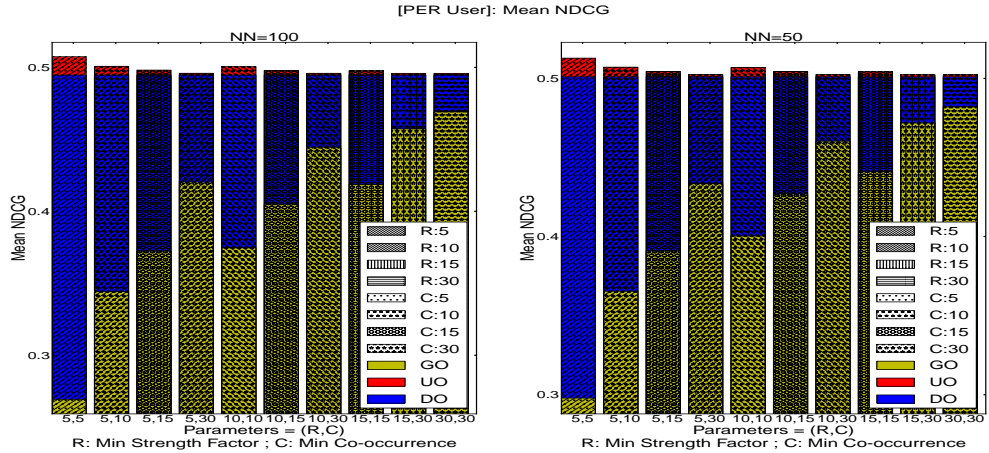}
  \caption{This figure shows the mean NDCG per \emph{user}, for 100 visual neighbors
  on the left, and 50 visual neighbors on the right. The performance is similar
  for both. Each bar on the x-axis corresponds to specific settings of the R and C
  parameters. The default ordering $DO$ is the baseline $XE$~\cite{li:11}}
  \label{fig:usr_ndcg_all}
 \end{figure*}

\begin{figure*}[ht!]
 \centering
  \includegraphics[height=.27\textheight,width=.9\textwidth]{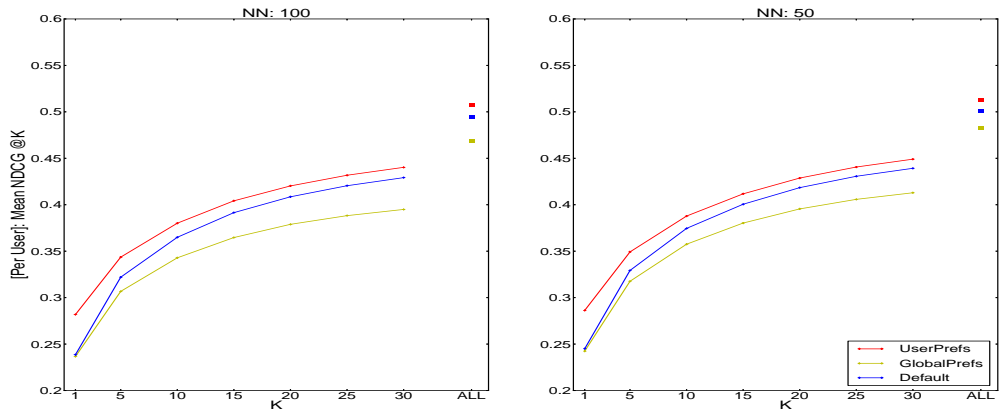}
  \caption{This figure shows the mean NDCG@K per \emph{user}, for 100 visual neighbors
  on the left, and 50 visual neighbors on the right. The performance is similar
  for both. The x-axis corresponds to K.}
  \label{fig:ndcg_k_usr}
 \end{figure*}

From \figurename~\ref{fig:usr_ndcg_all} and \ref{fig:ndcg_k_usr}, we notice that
our algorithm which personalizes via pairwise tag preferences (red bar/line)
always outperforms the state-of-the-art baseline \cite{li:11} (blue bar/line).

\figurename~\ref{fig:usr_ndcg_all} shows this increase in
performance for all parameters $R$ (the minimum strength factor described in 
section \ref{res:design_dec}), and $C$ (the minimum number of co-occurrences), 
that we considered. We notice that holding $C$ constant
as the parameter $R$ increases, the performance of our personalization method
decreases. This is likely due to a lack of enough data per user so that very
few tag pairs can meet our relatively aggressive over-fitting criteria. The
same trend is noticed when we hold $R$ constant and increase $C$.

We also evaluate the performance of the global (or ``average user'', yellow bar/line), pairwise
preferences, and we see that enforcing pairwise orders based on a global/average
notion of preference actually degrades the performance. Except for $K=1$,
the $nDCG@K$ metric for the global preferences is worse than the baseline.
This might be because, users tend to use more global/popular tags in the
beginning of their ranked lists, so that we are actually able to learn with
enough data, certain global biases among the popular tags. Beyond that, because
users tag differently, enforcing a global preference is actually counterproductive.
We also observe the reverse trend as we fix $C$ and vary $R$ (and vice-versa) to that
of the personal preferences. This is because to estimate global preferences
we need to observe a lot more data than for a single user, and so estimating
for pairs without a lot of co-occurrences will invariably lead to over-fitting.

\figurename~\ref{fig:ndcg_k_usr}, shows the 
$nDCG@K$ as $K$ varies under the best choices of parameters
$R$ and $C$, for both the global and personal preferences. The best choices
for the global were, $R=C=30$, and for the personal, $R=C=5$. We see, averaging
over both users and images(included in the appendix), that the baseline outperforms the global (except at
$K=1$), and the personal outperforms the baseline. These observations imply that
picking lower parameters is usually better for the personal preferences, while
higher values of $R$ and $C$ are necessary for the global preferences to be useful.

Table \ref{tab:avg_stats} shows that our method on average is $6\%$ better
than the baseline method, and this improvement goes up to about $30\%$,
for $nDCG@1$, as can be seen from Figure~\ref{fig:ndcg_k_usr}.

The above observations are true for both the average across all the images
in our test set, and across all the users' mean $nDCG$. This shows that the
gains we get are relatively consistent for each user. This combined with
the fact that the global pairwise preferences under-perform, imply that 
true personal pairwise preferences indeed exist, can be estimated, and can
be used to improve the performance of automatic image tagging. 

We include more detailed plots in the supplemntary appendix.


\section{Conclusion}

In this work we proposed a new measurement of tag preferences, and
demonstrated that there is indeed a tag-order bias, that is, when a user
mentions tag $a$ before tag $b$, in a list of tags for a given image, the
user is implying that he prefers, or considers $a$ to be of greater importance
than $b$. We showed that this bias can be learned from historical data using
the maximum likelihood estimate based on a pair's co-occurrence, and subsequently 
showed that such information can be exploited to improve the performance of 
current state-of-art automated image tagging systems. 

We also defined a new tagging objective function that assumes the inherent
pairwise bias between tags, and provided an algorithm which optimizes the
new objective (under some mild conditions), and helped verify our claims and
assumptions.

This leads us to conclude that although there are many visual factors
that may affect what tags a user will provide for an image, it is useful
to characterize instead (or rather in conjunction) the users' tagging
habits to learn what tags are of more importance to the users, whether
they are visually motivated or not, and automatic tagging systems should
employ this technique to improve their overall performance.


\section{Future Work}

We believe that there are several ways that this work could be extended or
extend other works. One direction we see is working on algorithms that 
provide tight guarantees for solving our tagging objective, or even solves it
optimally, even in the presence of dependent cycles. Another direction comes
from the observation that most tag pairs never occur, but it may be possible
to learn a function that maps a pair of tags to a real number indicating the 
direction and strength of the preference. It would also be interesting to 
see how the assumptions made in this paper can be used to extend the 
works on learning for personalized ranking \cite{rendle:09, rendle:10, pan:13}.
We think it would also be interesting to explore the cognitive dimensions that
drive tag ordering, and how these cognitive dimensions contribute to the tagging
choices both independently and collectively.

%
\bibliographystyle{IEEEbib}
\small{
\bibliography{icme16_bib}  
}

\end{document}